\documentclass[longbibliography,aps,prx,twocolumn,superscriptaddress,showkeys,amsfonts,amssymb,amsmath,floatfix]{revtex4-1}

\usepackage{graphicx}
\usepackage{dcolumn}
\usepackage{bm}
\usepackage{amsmath,amsbsy,amsfonts,revsymb}
\usepackage{float}
\usepackage{color}
\usepackage{inputenc}
\usepackage{physics}
\usepackage{setspace}
\usepackage{multirow}
\usepackage{array}
\usepackage{tabularx}
\usepackage{bm}
\usepackage[colorlinks=true,linkcolor=blue,citecolor=blue,urlcolor=blue]{hyperref}

\def\TTO/{$\mathrm{Tb_2Ti_2O_7}$}
\def\wn/{ cm$^{-1}$}

\def\edc/{\ensuremath{\mathbf{E}}}

\def\kvec/{\textbf{k}}
\def\Bfield/{\textbf{B}}
\def\Hfield/{\textbf{H}}
\def\Efield/{\textbf{E}}
\def\Hac/{\Hfield/$^\omega$}
\def\Eac/{\Efield/$^\omega$}

\graphicspath{{./pdf/}}

\begin{document}
	
	
	
	\title{Terahertz Magneto-optical investigation of quadrupolar spin-lattice effects in magnetically frustrated Tb$_2$Ti$_2$O$_7$}

	\author{K. Amelin}
	\affiliation{National Institute of Chemical Physics and Biophysics, Tallinn 12618, Estonia}
    \author {Y. Alexanian}
    \affiliation{Universit\'e Grenoble Alpes, CNRS, Institut N\'eel, 38000 Grenoble, France}
	\author{U. Nagel}
	\affiliation{National Institute of Chemical Physics and Biophysics, Tallinn 12618, Estonia}
    \author{T. R\~{o}\~{o}m}
    \affiliation{National Institute of Chemical Physics and Biophysics, Tallinn 12618, Estonia}
    \author{J. Robert}
    \affiliation{Universit\'e Grenoble Alpes, CNRS, Institut N\'eel, 38000 Grenoble, France}
    \author{J. Debray}
    \affiliation{Universit\'e Grenoble Alpes, CNRS, Institut N\'eel, 38000 Grenoble, France}
    \author{V. Simonet}
    \affiliation{Universit\'e Grenoble Alpes, CNRS, Institut N\'eel, 38000 Grenoble, France}
    \author{C. Decorse}
    \affiliation{ ICMMO, Universit\'e Paris-Saclay, CNRS, 91400 Orsay, France}
    \author{Z. Wang}
	\affiliation{Institute of Radiation Physics, Helmholtz-Zentrum Dresden-Rossendorf,01328 Dresden, Germany }
    \affiliation{Institute of Physics II, University of Cologne, 50937 Cologne, Germany}
    \author {R. Ballou}
    \affiliation{Universit\'e Grenoble Alpes, CNRS, Institut N\'eel, 38000 Grenoble, France}
    \author{E. Constable}
    \affiliation{Institute of Solid State Physics, Vienna University of Technology, 1040 Vienna, Austria}
    \author {S. de Brion}
    \affiliation{Universit\'e Grenoble Alpes, CNRS, Institut N\'eel, 38000 Grenoble, France}
    \email{sophie.debrion@neel.cnrs.fr}

\date{\today}
	
	\begin{abstract}
	Condensed matter magneto-optical investigations can be a powerful probe of a material's microscopic magneto-electric properties. This is because subtle interactions between electric and magnetic multipoles on a crystal lattice show up in predictable and testable ways in a material's optical response tensor, which dictates the polarization state and absorption spectrum of propagating electro-magnetic waves. Magneto-optical techniques are therefore strong complements to probes such as neutron scattering, particularly when spin-lattice coupling effects are present. Here
we perform a magneto-optical investigation of vibronic spin-lattice coupling in the magnetically frustrated pyrochlore \TTO/. Coupling of this nature involving quadrupolar mixing between the Tb$^{3+}$ electronic levels and phonons in \TTO/, has been a topic of debate for some time. This is particularly due to its implication for describing the exotic spin-liquid phase diagram of this highly debated system. A manifestation of this vibronic effect is observed as splitting of the ground and first excited crystal field doublets of the Tb$^{3+}$ electronic levels, providing a fine structure to the absorption spectra in the terahertz (THz) frequency range. In this investigation, we apply a static magnetic field along the cubic [111] direction while probing with linearly polarized THz radiation. Through the Zeeman effect, the magnetic field  enhances the splitting within the low-energy crystal field transitions revealing new details in our THz spectra. Complementary magneto-optical quantum calculations including quadrupolar terms show that indeed vibronic effects are required to describe our observations at 3 K. A further prediction of our theoretical model is the presence of a novel magneto-optical birefringence as a result of this vibronic process. Essentially, spin-lattice coupling within \TTO/ may break the optical isotropy of the cubic system, supporting two different electromagnetic wave propagations within the crystal. Together our results reveal the significance of considering quadrupolar spin-lattice effects when describing the spin-liquid ground state of \TTO/. They also highlight the potential for future magneto-optical investigations to probe complex materials where spin-lattice coupling is present and reveal new magneto-optical activity in the THz range.

	\end{abstract}
	

    \maketitle
	
\section{Introduction}	
	Interplay between spin and lattice degrees of freedom is the premise behind a range of intriguing phenomena in condensed matter systems. When we consider the fundamental role lattice geometry plays in the formation of conventional periodic magnetic order, this notion is perhaps unsurprising. Nevertheless,  when energetically favorable compensations between these degrees of freedom occur, we often find novel and potentially functional material properties emerge.
	 A case in point is found in the spin-Peirls transition of antiferromagnetic quantum spin chains where --- in order to lower the total energy of the system --- the lattice periodically contracts or dimerizes, thus favoring the formation of spin singlets, along with a global energy gap of their excitations \cite{Comes1973}.
	Another example is that of type II multiferroics, where the lattice reacts to a low-symmetry magnetic ordering by breaking its inversion symmetry and inducing a polar ferroelectric phase as a result of concomitant structural deformations \cite{Khomskii2009}. Spin-lattice effects are also present in magnetically frustrated systems where relaxations in the elastic degrees of freedom can lift the degeneracy of magnetic configurations promoting a long-range N\'{e}el order \cite{Lee2000, Tchernyshyov2002}. On the other hand, a dynamic interplay between the spins and lattice of a frustrated system can be perpetually destabilizing, inhibiting any type of order \cite{Gardner2010}. Indeed, this scenario seems to be the case in magnetically frustrated Tb$_2$Ti$_2$O$_7$, which fails to develop any long-range magnetic order or static frustrated configuration. Rather, a fluctuating spin liquid behavior is observed, persisting down to temperatures as low as 50\,mK \cite{Gardner2003}. A precise description of this peculiar magnetic ground state remains a hotly debated topic, although it is believed that spin-lattice effects play an important role \cite{Nakanishi2011, Bonville2011, Guitteny2013, Fenell2014}.
	
	\begin{figure}
    \resizebox{\columnwidth}{!}{\includegraphics{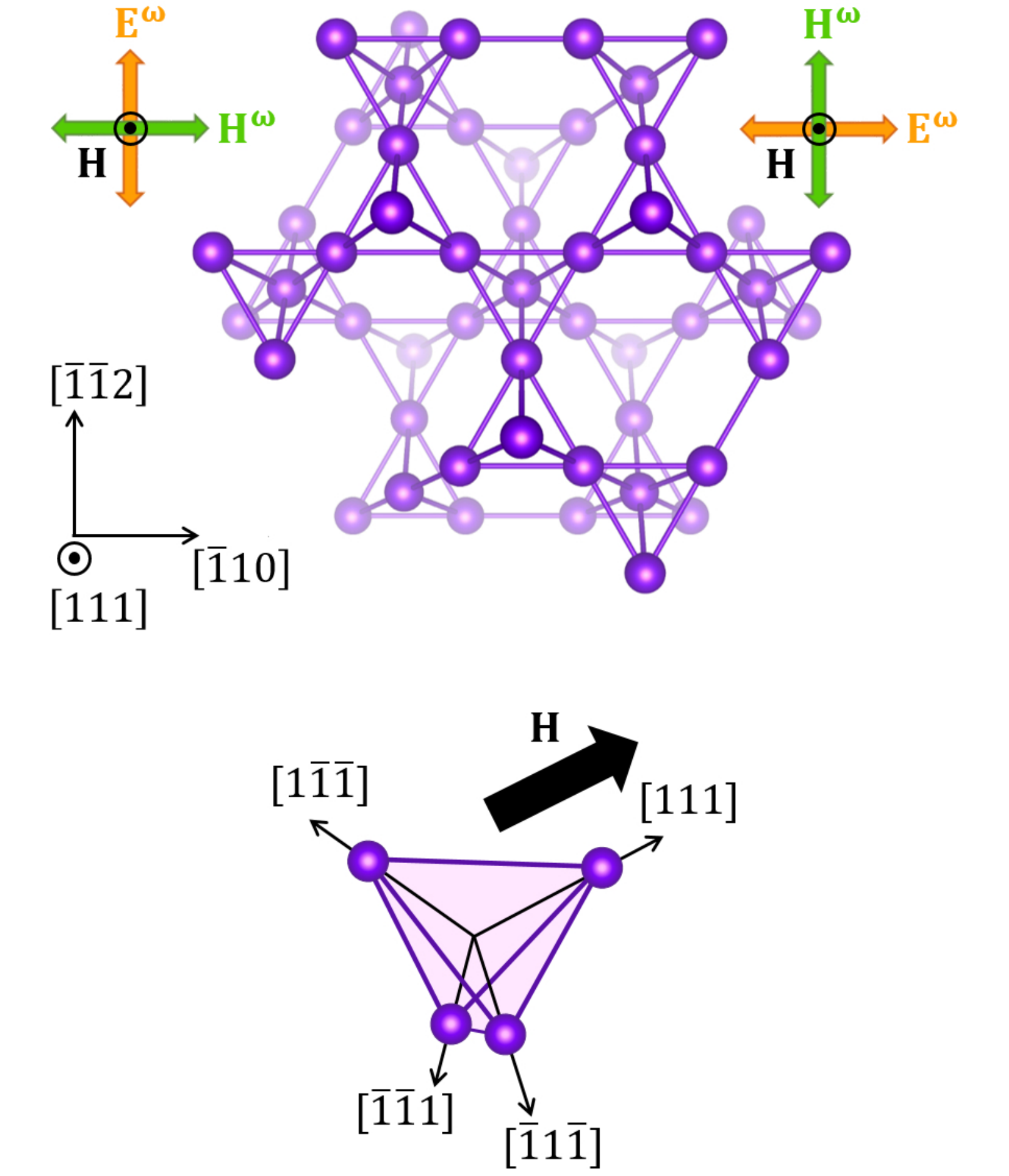}}
		{\caption{\TTO/ cubic structure with the Tb$^{3+}$ network. Upper part:  Cross-section viewed along the [111] direction showing the connection between tetrahedra. lower part: one single tetrahedron.  Two orthogonal linear polarizations of the incident light beam with $\kvec/\parallel [111]$ are shown in the upper corners, where the electric field vector \Eac/ and magnetic field vector \Hac/ oscillate parallel to $[\bar{1}\bar{1}2]$ or $[\bar{1}10]$. The static magnetic field H, shown in black, is applied perpendicular to the THz polarization plane, along one of the diagonals of the cube, the [111] direction.}}
    \label{fig:structure}
	\end{figure}

	In \TTO/, magnetic Tb$^{3+}$ ions are arranged in a network of corner-sharing tetrahedra, forming the so-called pyrochlore lattice shown in Fig. \ref{fig:structure}. Among rare-earth pyrochlores, \TTO/ is possibly the least understood, despite having been studied for over two decades. It certainly exhibits noticeable spin-lattice coupling effects. These are observable in X-ray diffraction experiments \cite{Ruff2010} and manifest as giant magnetostriction \cite{Aleksandrov1985}, elastic softening \cite{Nakanishi2011}, and  pressure-induced magnetic ordering \cite{Mirebeau2002}. More recently, inelastic neutron scattering \cite{Guitteny2013,Fennell2014,Ruminy2016} and THz spectroscopy \cite{Constable2017} measurements have highlighted the presence of vibronic coupling as a result of symmetry-allowed hybridization between phonons and the Tb$^{3+}$ crystal electric field (CF) states both within the ground and first excited doublets. In particular, these couplings involve quadrupolar operators that depend on the phonon mode inducing the local dynamical strains. Additionally, it is now well established that the phase diagram of \TTO/ is extremely sensitive to off-stoichiometry compositions \cite{ Taniguchi2013} and that Tb$_{2+x}$Ti$_{2-x}$O$_{7+y}$ enters a quadrupolar ordered phase below 500\,mK for x$\geq-0.0025$ \cite{Taniguchi2013}. Evidently, quadrupolar and spin-lattice effects play an important role in the ground state of \TTO/ and should be considered in any attempt to understand the spin-liquid behavior of this compound.
	
	In this study we focus on transitions between the low energy Tb$^{3+}$ CF excitations in \TTO/, performing magneto-optical observations of their modulation by an applied magnetic field. The first excited CF doublet is separated from the ground state doublet by $\Delta \approx 1.5$\,meV (0.37\,THz,12\,\wn/) \cite{Gingras2000,Gardner2001,Mirebeau2007}, and several other higher energy CF excitations also fall within the THz energy range \cite{Lummen2008}. To the best of our knowledge, no extensive magnetic-field dependence of the CF levels in \TTO/ has been previously performed. Our experimental results are compared to quantitative theoretical magneto-optical calculations incorporating a quantum mechanical vibronic coupling model. Magneto-optical studies of vibronic processes in magnetic molecules have a long history within the physical chemistry community \cite{Caldwell1969,Keiderling1981,Pawlikowski1984}. Within condensed matter physics, magneto-optical investigations are routinely applied to the study of  coupled dielectric and magnetic order parameters of multiferroics \cite{Kalashnikova2005,Pimenov2006,Miyahara2014}.
	Yet the combination of these ideas to probe novel spin-lattice effects in frustrated magnets has so far remained largely unexplored.
	The aim of this article is to provide better insight into the  magnetoelastic couplings and emerging hybrid excitations in frustrated \TTO/ using magneto-optical and quantitative theoretical techniques. Hence, we aim at broadening our knowledge on the microscopic mechanisms responsible for the spin liquid and quadrupolar phases it exhibits while highlighting the potential for further magneto-optical investigations of complex magnetic materials.

	\section{Experimental details}

	A large single crystal of \TTO/ was grown by the floating zone method using similar experimental parameters as in \cite{Petit2012}. A plaquette, 220\,$\mu$m thick and 4\,mm in diameter, was shaped with the [111] direction of the cubic pyrochlore lattice normal to the sample surface. A wedge with an angle of $\sim2^\circ$ was used to avoid interference fringes in the spectra. Another piece of the single crystal cut in close proximity to the plaquette was used for specific heat measurements. The specific heat data revealed a behavior similar to results published for a Tb$_{2+x}$Ti$_{2-x}$O$_{7}$ composition with $x=0.0025$ \cite{Taniguchi2013} --- quite close to the spin liquid phase, but with a quadrupolar ordering temperature of 400\,mK.

	Terahertz transmission magneto-optical measurements were performed by Fourier transform spectroscopy using a Martin-Puplett interferometer based at the National Institute of Chemical Physics and Biophysics in Tallinn. The  \TTO/  sample was mounted inside of a superconducting magnet within a liquid helium bath cryostat. The transmitted THz signal was detected by a sensitive Si bolometer cooled to 300\,mK using pumped $^3$He in a separate cryogenic closed circuit. The spectral bandwidth of the setup is 3 -- 200\wn/ (0.4 -- 25\,meV). The bandwidth was further limited to 80\wn/ due to strong sample absorption at high energies.
	
	The polarization of the incident THz radiation is controlled by an aluminum wire-grid polarizer in front of the sample.
	The spectra were measured in the Faraday configuration with a static magnetic field up to 15 T applied along the [111] axis and the wave vector \kvec/ parallel to the magnetic field vector \Hfield/.

	At each field value, the spectrum was measured with two orthogonal polarizations, where the oscillating electric and magnetic fields  \{\Eac/,\Hac/\} were either along $ \{[\bar{1}\bar{1}2], [\bar{1}10]\} $ or $ \{[\bar{1}10],[\bar{1}\bar{1}2]\} $, as shown in Fig.~\ref{fig:structure}.
	
	The spectral absorption $\alpha$ of a sample with thickness $d$ is determined by $\alpha = -(1/d)\ln[(1-R)^{-2}I/I_0]$ where $I_0$ is the incident light intensity, $I$ is the transmitted intensity at the detector, and $R$ is the reflection coefficient at the sample surface.
	To reveal excitations that have magnetic-field dependent energies and/or intensities, a differential absorption is calculated by 
	$\alpha_i - \alpha_{\mathrm{ref}} = -(1/d) \ln(I_i/I_{\mathrm{ref}})$. Here $I_i$ and $I_{\mathrm{ref}}$ are the transmitted light intensities detected at two different values of the magnetic field strength. Here, for $I_{\mathrm{ref}}$ we use a reference spectrum measured at 0\,T. The primary contribution to the reflection coefficient is the dielectric response of the phonon spectrum in the infrared range (100-1000cm-1) \cite {Constable2017}. We can then safely assume that the reflection coefficient is independent of the magnetic field strength. Therefore, the reflectivity in the differential absorption $\alpha_i - \alpha_{\mathrm{ref}}$ naturally cancels out in the THz range for Tb2Ti2O7. To deal with negative values in $\alpha_i - \alpha_{\mathrm{ref}}$ generated by spectral features in the reference spectrum ($\alpha_{\mathrm{ref}}$) that disappear under magnetic field, we subtract a statistically calculated baseline from all of the measured spectra. The baseline is created by taking the lowest value intensity at each frequency point from the set of measured spectra. Performing the baseline subtraction then corrects for any negative artifacts. The collection of baseline-corrected spectra together with the reference spectrum is what we define as the differential absorption $\Delta\alpha(H)$ that depends on the magnetic field strength.

	\section{THz spectroscopy results}
	
\begin{figure}
	
    \resizebox{\columnwidth}{!}{\includegraphics{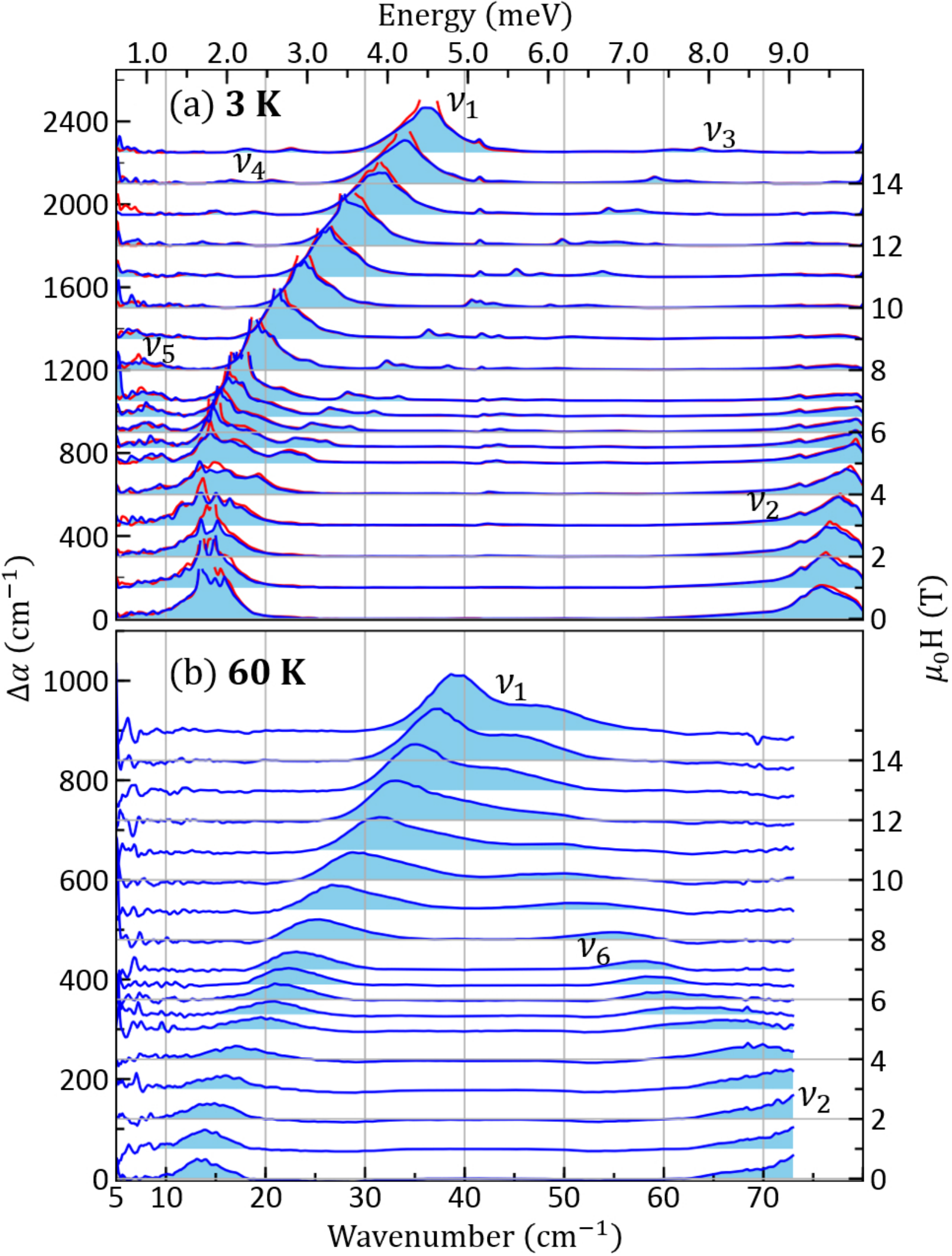}}
		\caption{$\Hfield/\parallel [111]$ magnetic-field dependence of the differential absorption $\Delta\alpha(H)$ in \TTO/  measured at (a) 3\,K and (b) 60\,K, with the reference absorption measured in zero field for two different THz polarizations (blue : \Hac/ along $[\bar{1}\bar{1}2]$, red: \Hac/ along $[\bar{1}10]$).
			The spectra are offset vertically in proportion to $H$. Shaded areas below the curves are included as a guide to the eye to highlight the different absorption bands.}
\label{fig:spectrum}
	\end{figure}

	The magnetic-field dependence of the differential absorption spectrum of \TTO/ is shown in Fig. \ref{fig:spectrum} at two different temperatures, 3\,K and 60\,K. The two different THz polarizations do not show any significant differences and are only plotted at 3\,K.
	A wide absorption band (designated $\nu_1$) is observed centered at 14\,\wn/, in agreement with previous THz studies \cite{Constable2017, Lummen2008}. It corresponds to the transition between the Tb$^{3+}$ ground state doublet and the first excited CF doublet. When the magnetic field is increased above 5\,T, the absorption band appears to broaden with a slight decrease in amplitude and a shift to higher energy. At approximately the same field value, weaker excitations emerge. Two of them ($\nu_3$ above 20\,\wn/ and $\nu_4$ around 10\,\wn/) harden with increasing magnetic field, while another one ($\nu_5$) softens and disappears below 5\,\wn/. Another broad absorption band ($\nu_2$) is seen at 75\,\wn/ at fields below 6\,T, which most likely corresponds to a transition from the ground-state level to the second excited CF level.
	
	At 60\,K, $\nu_1$ is still present but with a lower intensity due to thermal depopulation of the ground state. It splits into two different branches in high fields. On the other hand, $\nu_3$, $\nu_4$ and $\nu_5$ are no longer observed at 60\,K, while $\nu_2$ has a new component ($\nu_6$) that softens with magnetic field.
	
Combined intensity maps of the field dependence of the differential absorption are shown in Figs.~\ref{fig:calculations}a,b for the measurements at 3\,K and 60\,K, respectively. Together the results demonstrate a high degree of modulation in the CF energy-level scheme of Tb$^{3+}$ ions in \TTO/ within a magnetic field. In order to better understand this modulation and to determine the contribution from spin-lattice effects, we now turn to a comparison with theoretically calculated spectra. 

\section{Theoretical absorption calculations}
	
In order to understand the \TTO/ absorption spectra, we use linear response theory, where the sample response to the THz wave of angular frequency $\omega$ is described by the complex magnetic susceptibility tensor $[\chi(\omega)]$. Here only the magnetic part of the THz wave is considered. Omitting possible electric effects is valid here since all the relevant CF transitions occur within the first multiplet i.e. between states of same parity. Without vibronic coupling, the compound remains in the cubic symmetry and the susceptibility tensor is diagonal. The absorption of a propagating THz wave with wavevector $\mathbf{k}$ is then written \cite{Chaix2014}
\begin{equation}
\Delta\alpha\left(\omega\right) = 2k''\left(\omega\right)\approx \frac{n\omega}{c}\chi''\left(\omega\right)
\label{absorption_isotrope}
\end{equation}
as derived by solving Maxwell equations in an isotropic medium with weak dissipation. Here $c$ is the speed of light in vacuum, $n$ is the  refractive index of the medium and $k$ is the wavenumber. Here and further, the notations prime and double prime refer respectively to the real and imaginary part of a quantity. The wavevector can be written as $\mathbf{k} = k\mathbf{u}$ where $\mathbf{u}$ is a vector perpendicular to the wavefront. In the isotropic case that includes the cubic symmetry relevant to pyrochlore compounds, the Poynting vector of the electromagnetic wave $\mathbf{S}=\textrm{Re}\left[\mathbf{E}^{\omega}\times\overline{\mathbf{H}^{\omega}}\right]$ is collinear with $\mathbf{k}$ outside and inside the material. Here $\overline{\mathbf{H}}$ stands for the complex conjugate of $\mathbf{H}^{\omega}$. The refractive index $n$ is considered constant since the main contribution comes from optical phonons that are at energies higher than the measured THz range (see Ref. \cite{Constable2017} and supplementary material therein). This is generally the case in oxides below 80\,\wn/ where absorption is low and very few phonons are present. In our calculations we used $n=7.7$ as deduced from the dielectric constant of \TTO/ at 6\,K \cite{Constable2017}.

\begin{figure}
		 \resizebox{\columnwidth}{!}{\includegraphics{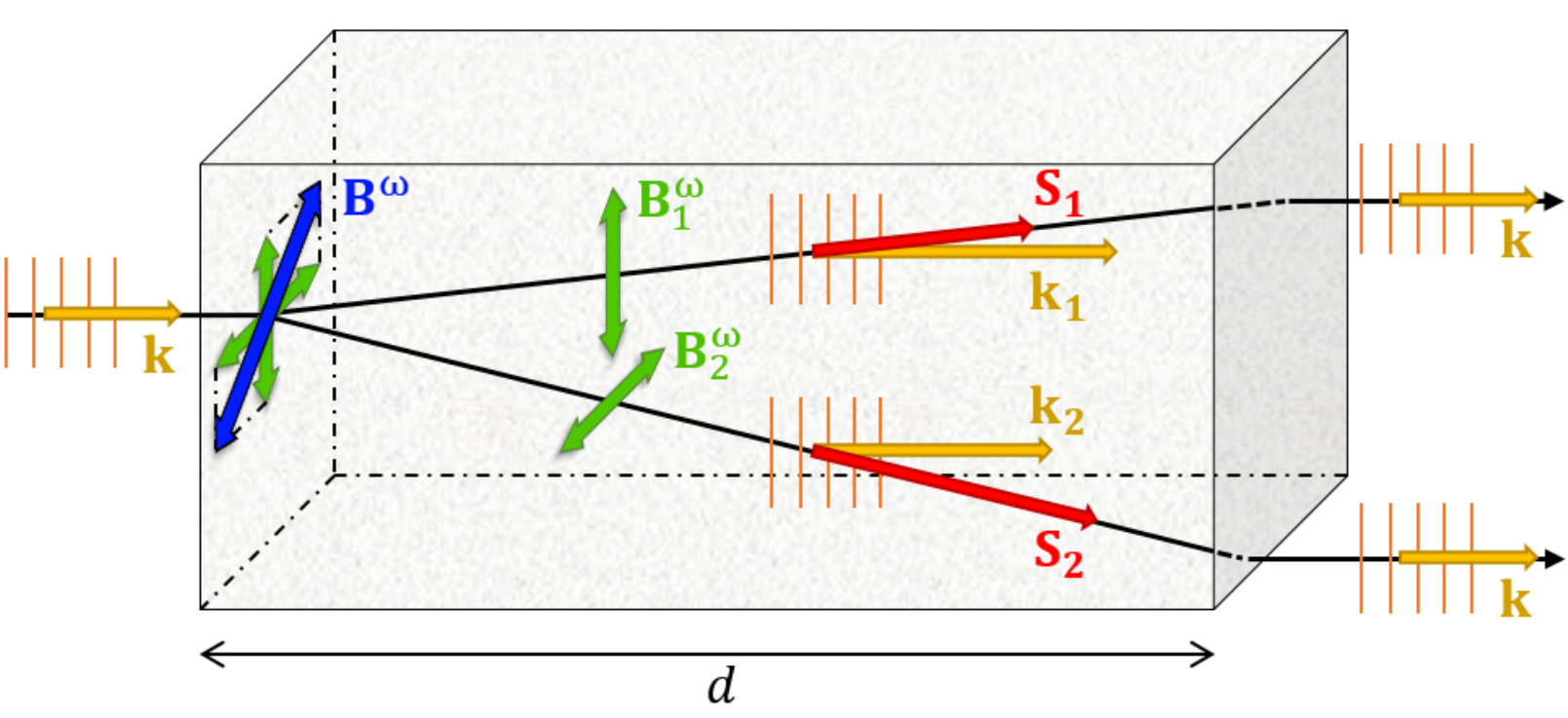}}
		\caption{ THz wave propagation in an anisotropic medium. The wave, linearly polarized along $\mathbf{B}^{\omega}$ (blue), is decomposed into the two --- orthogonal and linearly polarized for simplicity of the picture --- normal modes polarized along $\mathbf{B_{1}^{\omega}}$ and $\mathbf{B_{2}^{\omega}}$ (green). Inside the crystal the two waves are normal modes, and the rays propagate independently in the direction of their Poynting vector $\mathbf{S_{1}}$ and $\mathbf{S_{2}}$ (green) with  wavevector $\mathbf{k_{1}}$ and $\mathbf{k_{2}}$ respectively, which have the same direction (orange). At the output face, in vacuum, the Poynting vectors are collinear again.}
    \label{fig:Maxwell}
	\end{figure}

We now introduce vibronic couplings, which arise from dynamical strains that break the local symmetry. Thus, the four Tb$^{3+}$ sites of a tetrahedron become inequivalent and the whole tetrahedron has to be considered. At this scale, the magnetic susceptibility tensor remains diagonal but becomes slightly anisotropic, quite similarly to birefringent crystals in optics. When a static magnetic field is applied along the $[111]$ cubic direction, non-diagonal components appear in the susceptibility tensor. Two normal modes (indexed by $\alpha=\left\{1,2\right\}$ with different absorption are then derived from the Maxwell's equations.  They are characterized by their wavevectors $\mathbf{k}_{\alpha}$ and their Poynting vectors $\mathbf{S_{\alpha}} = \textrm{Re}\left[\mathbf{E}_{\alpha}^{\omega}\times\overline{\mathbf{H}_{\alpha}^{\omega}}\right]$ that are no longer collinear. The THz wave polarisation in the material characterized by the magnetic induction $\mathbf{B}^{\omega} = \mu_{0}\left(1 + [\chi(\omega)]\right)\mathbf{H}^{\omega}$ is no longer collinear with $\mathbf{H}^{\omega}$. This is illustrated in Fig.~\ref{fig:Maxwell}. A similar effect has been predicted by considering electric quadrupole and magnetic dipole mixing in antiferromagnets \cite{Graham1992}. To our knowledge, this example involving vibronic processes in a frustrated magnet has not been previously reported.

The total absorption in the crystal will then contain contributions of these two modes. The transmitted intensity is given by
\begin{equation}
\begin{aligned}
I &= \textrm{e}^{-2k''_{1}l}\textrm{Re}\left[\mathbf{s_{11}}\right] + \textrm{e}^{-2k''_{2}l}\textrm{Re}\left[\mathbf{s_{22}}\right] \\
&+ \textrm{e}^{-(k''_{1}+k''_{2})l} \textrm{Re}[\textrm{e}^{i\Delta k'l}\mathbf{s_{12}}+\textrm{e}^{-i\Delta k'l}\mathbf{s_{21}}],
\label{Cas_general}
\end{aligned}
\end{equation}
where $\Delta k' = k'_{1}-k'_{2}$ and  $\mathbf{s_{ij}}\propto \mathbf{E}_{i}^{\omega}\times\overline{\mathbf{H}_{j}^{\omega}}$. The last term in Eq. \ref{Cas_general} is similar to an interference term when the two normal modes are not orthogonal. The case of orthogonal modes has been developed in Ref.~\cite{Saleh1991} and the associated absorption is written as:
\begin{equation}
\Delta\alpha\left(\omega\right) = \frac{2k''_{1}\left(\omega\right)\textrm{Re}\left[\mathbf{s_{11}}\right] + 2k''_{2}\left(\omega\right)\textrm{Re}\left[\mathbf{s_{22}}\right]}{\textrm{Re}\left[\mathbf{s_{11}}\right] + \textrm{Re}\left[\mathbf{s_{22}}\right]}.
\label{Modes_orthogonaux}
\end{equation}
When $k_{1} = k_{2} \equiv k$, the isotropic case (equation \ref{absorption_isotrope}) is recovered.

\begin{figure*}
		\resizebox{17cm}{!}	{\includegraphics{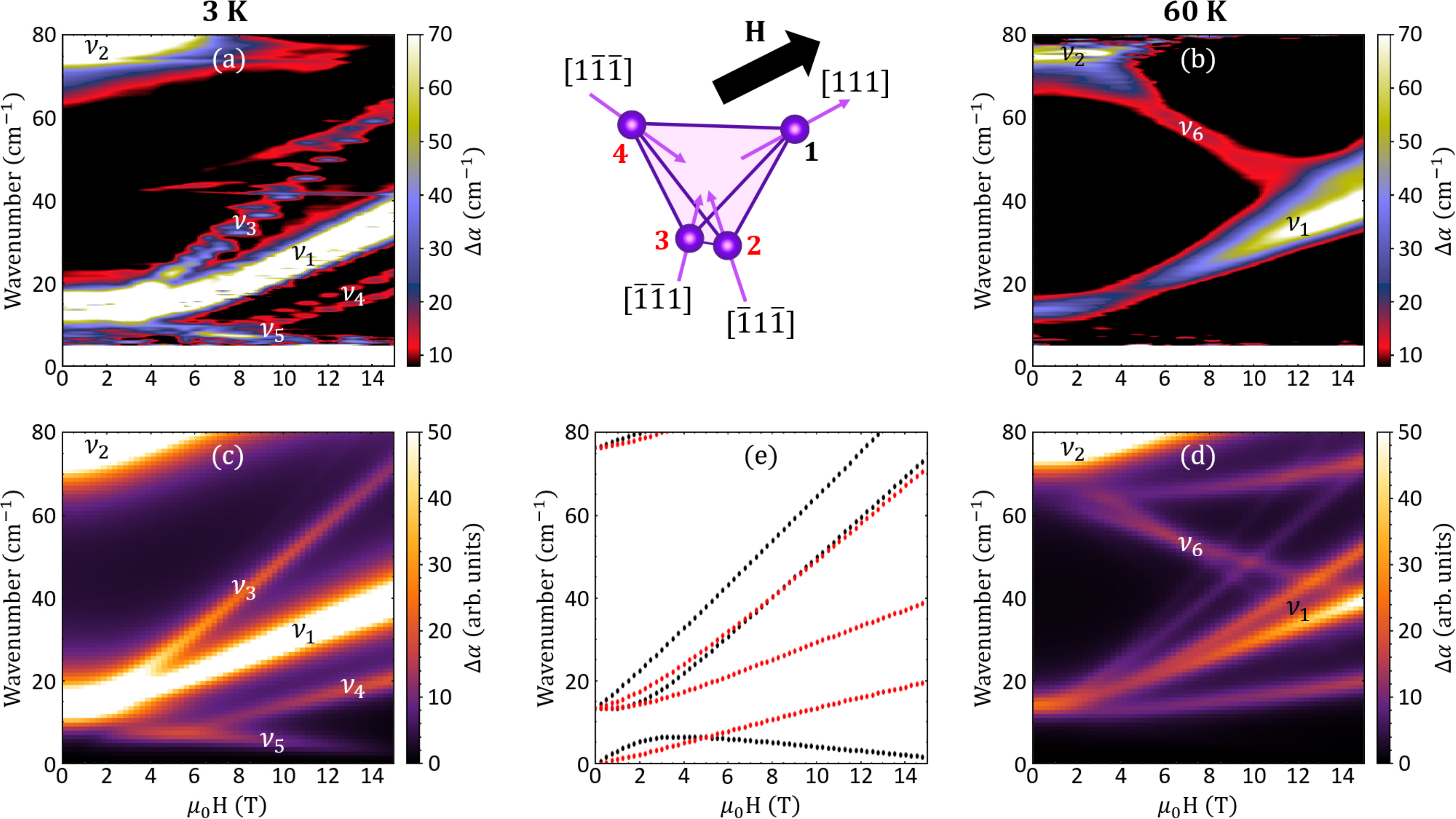}}
		\caption{Experimental and calculated THz absorption as a function of magnetic field applied along the [111] direction of \TTO/. The panels show the experimental results for 3\,K (a) and 60\,K (b), and the theoretical calculations for 3\,K (c) and 60\,K (d). The four Tb$^{3+}$ sites on the tetrahedron (one with the field along its 3-fold axis shown in black and the three remaining sites shown in red) and the corresponding field-dependence of the calculated energy of their absorption branches at 3\,K are presented in the middle panels (e). The different observed branches are labeled $\nu_1$ to $\nu_6$.}
\label{fig:calculations}
	\end{figure*}

Equation \ref{Modes_orthogonaux} allows us to calculate the differential absorption for the wavevector of the two normal modes $k_{i}$. These are functions of the complex magnetic susceptibility tensor components $\chi_{ij}$ which are given by:
\begin{equation}
\begin{aligned}
\chi_{ij}\left(\omega\right) = \frac{\mu_{0}\left(g_{J}\mu_{B}\right)^{2}}{V}\sum_{mn}\frac{P_{n}-P_{m}}{\left(E_{m}-E_{n}-\hbar\omega\right)^{2}+\Gamma^{\textrm{ }2}}\\
\left[i\Gamma + \left(E_{m}-E_{n}-\hbar\omega\right)\right]J^{i}_{nm}J^{j}_{mn},
\end{aligned}
\end{equation}
where $g_{J}$ is the Land$\acute{e}$ factor, $V$ is the sample volume, $E_{m,n}$ is the energy of the different electronic levels,  $P_n$ and $P_m$ are the thermal populations of initial and final states, and $J^{i}_{nm}$ is the matrix element between electronic states $|n\rangle $ and $|m\rangle$  of their  angular momentum  in the $i$ direction. One single linewidth $\Gamma$ is used for simplicity.
The energy levels  are determined by diagonalizing the corresponding Hamiltonian. We will consider only those levels that fall within the THz energy range at low temperatures --- i.e. the ground, first and second levels.
The Hamiltonian consists of several terms. The first one, the CF Hamiltonian, describes the effects of the charges surrounding each Tb$^{3+}$ ion on its electronic states in its local D$_{3d}$ symmetry. These ions generate four Bravais lattices from each of the four vertices of an initial Tb tetrahedron, the basic element of the pyrochlore structure. The axis of the 3-fold symmetry for each ion is parallel to a distinct member of the family of $\left<\mathrm{111}\right>$ diagonals of the cubic structure characterizing the global symmetry of the material. By selecting this local 3-fold axis as the quantization z axis and the local 2-fold axis as the x- axis which gives rise to the point group D$_{3d}$, the CF Hamiltonian is written for each ion in the same form
\begin{equation}
\widehat{\mathcal{H}}_{CF} = B_{2}^{0}\widehat{\mathcal{O}}_{2}^{0} + B_{4}^{0}\widehat{\mathcal{O}}_{4}^{0} + B_{4}^{3}\widehat{\mathcal{O}}_{4}^{3} + B_{6}^{0}\widehat{\mathcal{O}}_{6}^{0} + B_{6}^{3}\widehat{\mathcal{O}}_{6}^{3} + B_{6}^{6}\widehat{\mathcal{O}}_{6}^{6},
 \label{H_CEF}
\end{equation}
where the expansion in Stevens operators (quadrupolar $\widehat{\mathcal{O}}_{2}^{0}$, hexadecapolar $\widehat{\mathcal{O}}_{4}^{0}$, $\widehat{\mathcal{O}}_{4}^{3}$, and hexacontatetrapolar  $\widehat{\mathcal{O}}_{6}^{0}$, $\widehat{\mathcal{O}}_{6}^{6}$) terms is given by the local D$_{3d}$ symmetry of the Tb$^{3+}$ ions. For correspondence with Wybourne and angular momentum operators, see Appendix \ref{appendixI} and \ref{appendixII}.

When a static magnetic field $H$ is applied along the [111] direction of the pyrochlore cubic lattice, one Tb$^{3+}$ ion out of four has its 3-fold axis along the magnetic field, while the three remaining sites have their 3-fold axes at the same colatitude (polar angle) from the magnetic field and behave similarly. The corresponding Zeeman Hamiltonian is given by
\begin{equation}
\widehat{\mathcal{H}}_{Z} = -g_{J}\mu_{B}\mu_{0}\bf{H}\cdot\mathbf{\widehat{J}},
\end{equation}
where $g_{J}\mu_{B}\mathbf{\widehat{J}}$ is the Tb$^{3+}$ ion's total magnetic moment ($J=6$).

Finally, the total Hamiltonian for non interacting tetrahedra is given by
\begin{equation}
\widehat{\mathcal{H}} = \sum_{k=1,4}\widehat{\mathcal{H}}_{CF}^k  + \widehat{\mathcal{H}}_{Z}^k.
\end{equation}

As shown in Eq.~5, the CF Hamiltonian is described in a local frame for each Tb ion, while the Zeeman term is better described in the global cubic frame. Therefore, the Stevens operators must be rotated from the local frame associated to each Tb ion, to the global cubic frame.

The results of the calculations using the Hamiltonian of Eq. 4 (without vibronic coupling) are shown in Fig.~\ref{fig:calculations}. The CF parameters were chosen from the literature \cite{Ruminy2016} except for $B_2^0$ and $B_2^4$, which were slightly adjusted to match the 14\,\wn/  transition observed at 3\,K and 0\,T (see table \ref{tab:CFparameters} and Appendix \ref{appendixII}). The wave function for the ground and first excited doublets are given in Appendix \ref{appendixIII}. An effective Land\'{e} factor $g_J \approx 1.4$ is  deduced from the magnetic-field dependence of the measured spectra. The obtained value is slightly lower than $g_J = 1.5$ expected for a pure Tb$^{3+}$ ion ground multiplet and reveals the $J-$mixing effects seen in the intermediate coupling regime \cite{Ruminy2016}. A single line width of 2.4\,\wn/ is used, in agreement with the zero field data. We find no significant dependence on the polarization of the THz radiation in the calculated spectra, consistent with experiment.
The calculated absorption has two contributions when the applied static magnetic field is varied (see panel(e) in Fig.~\ref{fig:calculations}): one  from the Tb$^{3+}$ site (1) that has its local 3-fold axis along the magnetic field direction, the other one from the 3 others sites (2-4) on the tetrahedron that have their local 3-fold axes at 109.5\,degrees relative to the applied magnetic field. When the magnetic field is increased, the ground and first excited CF levels --- both of which are doublets --- split into two branches: a lower (and a higher) frequency branch that decreases (respectively increases) in energy with increased magnetic field. The second CF level is a singlet and its energy increases with the magnetic field.

As seen in Fig.~\ref{fig:calculations}, the agreement with the experimental data is already remarkable. The field dependence of the main excitations $\nu_1$ and $\nu_2$ is well reproduced at 3\,K. The first one originates entirely from sites 2-4 and corresponds to the transition to their first CF level. The second one has contributions from all sites and is the transition to the second CF level.  Weaker features in the absorption maps are also reasonably well reproduced: $\nu_3$ for the transition to the first CF level (upper branch for sites 2-4 and lower branch for site 1) and $\nu_4$ for the transition within the initial ground doublet for sites 2-4. Also note the peculiar magnetic field dependence of $\nu_5$: for a field lower than 3\,T it is the equivalent of $\nu_4$ for site (1), but above 3\,T, the transition to the first excited CF level occurs, producing the only excitation decreasing in energy with the magnetic field.
A third branch starting at 14\wn/ and increasing more rapidly under magnetic field than the other branches (visible in Fig.~\ref{fig:calculations}e) is calculated to be very weak in intensity. It is not visible in either of the calculated or measured absorption maps (Fig.~\ref{fig:calculations}c,a).

With our theoretical basis for the field dependence of the CF energy scheme we also reproduce the 60\,K results:  the main branches,  $\nu_1$ at 14\,\wn/  and $\nu_2$ at 70\,\wn/, as well as a new branch decreasing from 75\,\wn/ ($\nu_6$). Noticeably, two additional weak and rather flat branches are calculated around 65\,\wn/ and 14\,\wn/ but are not observed in the THz absorption spectra.

\begin{table}[!htbp]
 \label{tab:CFparameters}
 \begin{ruledtabular}
 \begin{tabular} {ccc}
  $B_{k}^{q}$& meV & K \\
\noalign{\vskip 0.5mm}  \hline \noalign{\vskip 0.5mm}
 $B_{2}^{0}$ & $-0.26$ & $-3.0$ \\
 $B_{4}^{0}$ & $4.5\cdot 10^{-3}$ & $5.2\cdot 10^{-2}$\\
 $B_{4}^{3}$ & $-4.1\cdot 10^{-2}$ & $-4.8\cdot 10^{-1}$ \\
 $B_{6}^{0}$ & $-4.5\cdot 10^{-6}$ & $-5.2\cdot 10^{-5}$\\
 $B_{6}^{3}$ & $-1.2\cdot 10^{-4}$ & $-1.4\cdot 10^{-3}$ \\
 $B_{6}^{6}$ & $-1.4\cdot 10^{-4}$ & $-1.6\cdot 10^{-3}$ \\
 \end{tabular}
\end{ruledtabular}
\caption{CF parameters used in the CF Hamiltonian.}
\end{table}

\begin{figure*}
		\resizebox{17cm}{!}	{\includegraphics{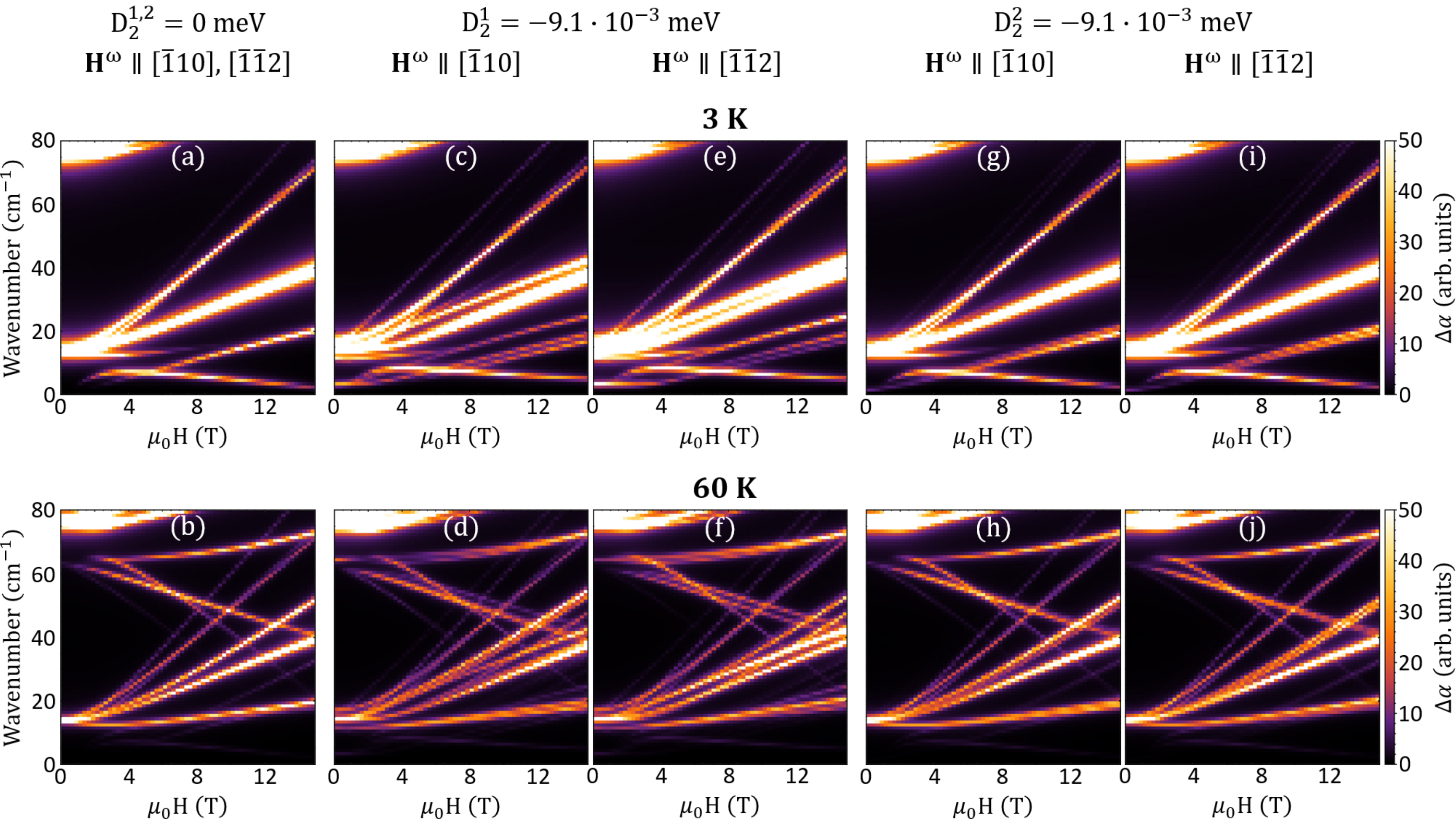}}
		\caption{Calculated absorption as a function of magnetic field applied along [111] at 3\,K and 60\,K for two THz polarizations with no vibron (a,b), with a vibron associated with $\mathcal{O}_{2}^{1}$ (c-f) and a vibron associated with $\mathcal{O}_{2}^{2}$ (g-j) quadrupolar operators.}
\label{fig:calculations_vibron}
	\end{figure*}

As the next step we have performed calculations including spin-lattice effects through vibronic couplings between the Tb$^{3+}$ crystal field excitations and transverse phonon modes. It has been shown that there are two vibronic processes present: one that couples the first excited Tb$^{3+}$ CF level with a silent optical phonon of $T_{2u}$ symmetry, and another one that involves an acoustic phonon coupled to both the ground and first excited CF levels \cite{Constable2017}. In particular, these spin-lattice couplings were shown to involve the Tb$^{3+}$ quadrupolar degrees of freedom, and give rise to the following symmetry-constrained vibronic Hamiltonian
 \begin{equation}
 \widehat{\mathcal{H}}_{vib} = D_{2}^{0}\widehat{\mathcal{O}}_{2}^{0} + D_{2}^{1}\left(\widehat{\mathcal{O}}_{2}^{1} + \widehat{\mathcal{O}}_{2}^{-1}\right) + D_{2}^{2}\left(\widehat{\mathcal{O}}_{2}^{2} + \widehat{\mathcal{O}}_{2}^{-2}\right)
 \label{H_vibron}
\end{equation}
when the vibronic coupling is assumed isotropic in the plane perpendicular to the 3-fold axis.
Note that the quadrupolar operator $\widehat{\mathcal{O}}_{2}^{0}$ is already present in the CF Hamiltonian. It accounts for the coupling to the silent optical phonon and will not change the symmetry of the system, but will simply renormalize its energy eigenvalues. On the other hand, $\widehat{\mathcal{O}}_{2}^{m}$ operators with $m=\pm1,\pm2$, associated with the acoustic phonons, are not present in the CF Hamiltonian. They induce a splitting of the ground and first excited CF doublets as described in \cite{Constable2017}. The associated wave functions are given in Appendix \ref{appendixIII}.  The resulting susceptibility tensor is no longer diagonal and the crystal becomes slightly birefringent.

The influence of both terms ($D_{2}^{1}$ and $D_{2}^{2}$) of the acoustical vibronic coupling on the calculated absorption spectra is presented in Fig.~\ref{fig:calculations_vibron}. A smaller line width of 0.5\,\wn/ was used in the calculation to better distinguish the different branches that appear due to the vibronic coupling. The Tb$^{3+}$ sites 2-4 are no longer equivalent and the associated branches are split into two or three components that are more or less distinguishable. This is particularly true for the lower-energy branch at 3\,K just below $\nu_1$,  where two groups of lines are now clearly observed in agreement with the experimental data for $\nu_4$. At 60\,K absorption branches are more spread out and therefore less intense. We suspect this could be an explanation for the absence of the flatter bands in the experimental data. Their combination of weak intensity and moderate field dependence would cancel them out in our background subtraction analysis method. Furthermore, we also note the slight polarization dependence that shows up at both temperatures.

\section{Discussion}

A comparison of the experimental and the simulated magneto-optical THz spectra demonstrates that the low-energy dynamics of \TTO/ is well captured by a simple Hamiltonian with only CF and Zeeman contributions. However, our results also indicate that including spin-lattice effects by way of vibronic coupling provides an improved agreement between theory and experiment with the addition of several weaker branches at both 3 K and 60 K. Although subtle, these features are clearly evident in the experimental data, particularly for the transition assigned $\nu_4$. They provide strong evidence that spin-lattice coupling is at play within the energy and time scales relevant to the ground state of this quantum spin liquid. In particular, the vibronic couplings of quadrupolar origin will lower the symmetry of the dynamical susceptibility tensor, which, without external magnetic field, becomes slightly orthorhombic. This implies a dynamic modulation of the local CF environment that describes the magnetic behavior including the potential for entanglement between the different CF levels.

Our results confirm that entanglement between the ground and first excited CF states through the vibronic process is of utmost importance in understating the phase diagram of \TTO/ at lower temperatures, as was already suggested in Ref. \cite{Constable2017}. It is difficult to unambiguously quantify the strength of the quadrupolar couplings associated to $\mathcal{O}_{2}^{1}$ and $\mathcal{O}_{2}^{2}$. However, according to the observed splittings of the different branches, the $D_{2}^{1}$ and/or $D_{2}^{2}$ terms fall within the energy range 0-2 meV.  Note that these operators, having E$_g$ symmetry in the local D$_{3d}$ environment, have E$_g\oplus$T$_{1g}\oplus$T$_{2g}$ symmetry in the global cubic environment. From a symmetry point of view, they are equivalent to a combination of tetragonal and trigonal stress. Nothing clearly distinguishes one operator from the other, except for their strength. As can be seen in Fig.~\ref{fig:calculations_vibron}(c,e,g and h), operator $\mathcal{O}_{2}^{1}$ has a larger effect on the splitting of the different branches than $\mathcal{O}_{2}^{2}$ does. As a matter of fact, the matrix elements of $\mathcal{O}_2^1$ between the ground and first excited states are five times larger than those of $\mathcal{O}_2^1$ at zero applied magnetic field, which would imply five times larger vibronic effects for equal $D_2^1$ and $D_2^2$ parameters. This is consistent with the oscillator strength associated with each of the quadrupolar operators, which depend on the structure of the ground and excited doublet states. Note that all of these operators act on the transverse components of the Tb$^{3+}$ angular momentum. It is then possible that these vibronic couplings have some role to play in the low temperature phase diagram of Tb$_{2+x}$Ti$_{2-x}$O$_{7+y}$ below 1 K, where a spin liquid or a quadrupolar ordered phase is observed \cite{Takatsu2016}.

Finally, due to the high magnetic fields used in this study, we have been able to refine with greater precision the crystal field parameters (see Table \ref{tab:CFparameters}) and the Land\'{e} factor $g_J \approx 1.4$. Within the energy range probed, there is no sign of spin waves down to 3K, which could be present due to a possible ordered magnetic state as observed by neutron diffraction at 40 mK \cite{Sazonov2013}.  Indeed our analysis is in perfect agreement with the measured "3-in /1-out, 3-out/1-in" spin orientation per tetrahedron  induced by  the  magnetic field applied along [111] as well as  the proposed dynamical Jahn-Teller model \cite{Sazonov2013}. Our results and analysis allow us to give a more precise description of these spin-lattice couplings.

\section{Conclusion}
Performing magneto-optical THz spectroscopy measurements of \TTO/, we have shown the magnetic field dependent evolution of the low energy CF level scheme for Tb$^{3+}$.
Using a simple model Hamiltonian incorporating CF and Zeeman contributions, the overall field dependent trends observed in our experiments are well reproduced by the theory, in particular showing multiple excitation branches that can be attributed to transitions between the different levels for each site in the elementary tetrahedra.
However, fFiner structure observed in the experiment cannot be captured by the simple model and is only reproduced after the inclusion of a vibronic spin-lattice coupling process where the ground and first excited CF doublets are hybridized with acoustic phonons by way of quadrupolar Stevens operators.
The results add further support to the growing evidence that spin-lattice coupling and quadrupolar terms are important when describing the frustrated ground state of \TTO/, a topic that is still under debate. Finally, we also predict that under an external magnetic field, these couplings induce a novel birefringent response of this otherwise cubic pyrochlore. While this effect has not been tested, a direct measurement would provide further support to the vibronic model. We suggest that this highlights the potential for future magneto-optical investigations aimed at probing complex magnetic phases where spin and lattice degrees of freedom are present. It also open new routes to design magneto-optically active materials.

	\begin{acknowledgments}		
		K.A., T.R. and U.N. acknowledge support of the Estonian Ministry of Education and Research with institutional research funding IUT23-3, and the European Regional Development Fund Project No. TK134.
		E.C. acknowledges the Austrian Science Foundation (FWF) for financial support of the EMERGE project (number P32404-N27).
		SciPy library \cite{Virtanen2020} for Python was used for the data analysis and representation, and the crystal structure was modeled in Vesta software \cite{Momma2011}.
	\end{acknowledgments}

K.A. and Y.A. contributed equally to this work. K.A.,T.R. and U.N. performed the THz measurements, K.A. analysed the experimental data. Y.A. developed the THz calculations with inputs from R.B., J.R., V.S. and S.deB. C.D. grew the single crystal, Y.A. and J.D. prepared the plaquettes. Z.W., E.C. and S.deB. have coordinated the project. K.A., Y.A., E.C. and S.deB. prepared the figures and wrote the paper with inputs from all authors.

\appendix
\section{Stevens equivalent operators}
\label{appendixI}

The Stevens equivalent operators $\widehat{\mathcal{O}}_{n}^{m}$ used in equations  \ref{H_CEF} and  \ref{H_vibron} can be expressed as a function of the angular momentum operators $\widehat{J}_{x,y,z}$ and $\widehat{J}_{+,-}$ of the rare earth ground multiplet. We give here their correspondence together with the one for Stevens operators using the x,y,z, notation. These operators are tabulated in references \cite{Stevens1952,Hutchings1964}.
We will use :
\begin{equation*}
\widehat{X} = J(J+1)\widehat{I}
\end{equation*}
where $\widehat{I}$ is the identity operator.  It follows:
\begin{equation*}
\widehat{\mathcal{O}}_{2}^{0} = \widehat{\mathcal{O}}_{z^{2}} = 3\widehat{J}_{z}- \widehat{X}
\end{equation*}
\begin{equation*}
\widehat{\mathcal{O}}_{2}^{1} = \widehat{\mathcal{O}}_{xz} = \frac{1}{2}\left(\widehat{J}_{z}\widehat{J}_{x}+\widehat{J}_{x}\widehat{J}_{z}\right)
\end{equation*}
\begin{equation*}
\widehat{\mathcal{O}}_{2}^{-1} = \widehat{\mathcal{O}}_{yz} = \frac{1}{2}\left(\widehat{J}_{z}\widehat{J}_{y}+\widehat{J}_{y}\widehat{J}_{z}\right)
\end{equation*}
\begin{equation*}
\widehat{\mathcal{O}}_{2}^{2} = 2\widehat{\mathcal{O}}_{x^{2}-y^{2}} =  \frac{1}{2}\left(\widehat{J}_{+}^{2}+\widehat{J}_{-}^{2}\right) = \widehat{J}_{x}^{2}-\widehat{J}_{y}^{2}
\end{equation*}
\begin{equation*}
\widehat{\mathcal{O}}_{2}^{-2} = 2\widehat{\mathcal{O}}_{xy} =  -\frac{i}{2}\left(\widehat{J}_{+}^{2}-\widehat{J}_{-}^{2}\right)= \widehat{J}_{x}\widehat{J}_{y}+\widehat{J}_{y}\widehat{J}_{x}
\end{equation*}

\begin{equation*}
\widehat{\mathcal{O}}_{4}^{0} = 35\widehat{J}_{z}^{4}-\left[30\widehat{X}-25\widehat{I}\right]\widehat{J}_{z}^{2} +  \left[3\widehat{X}^{2}-6\widehat{X}\right]
\end{equation*}
\begin{equation*}
\widehat{\mathcal{O}}_{4}^{3} = \frac{1}{4}\left[\widehat{J}_{z}\left(\widehat{J}_{+}^{2}+\widehat{J}_{-}^{2}\right)+\left(\widehat{J}_{+}^{2}+\widehat{J}_{-}^{2}\right)\widehat{J}_{z}\right]
\end{equation*}
\begin{equation*}
\widehat{\mathcal{O}}_{4}^{4} = \frac{1}{2}\left(\widehat{J}_{+}^{4}+\widehat{J}_{-}^{4}\right)
\end{equation*}

\begin{equation*}
\begin{split}
\widehat{\mathcal{O}}_{6}^{0} &= 231\widehat{J}_{z}^{6}-\left[315\widehat{X}-735\widehat{I}\right]\widehat{J}_{z}^{4}\\
&+ \left[105\widehat{X}^{2}-525\widehat{X}+294\widehat{I}\right]\widehat{J}_{z}^{2}\\
&-\left[5\widehat{X}^{3}-40\widehat{X}^{2}+60\widehat{X}\right]
\end{split}
\end{equation*}
\begin{equation*}
\begin{split}
\widehat{\mathcal{O}}_{6}^{3} &=\frac{1}{4}\left[\left\{11\widehat{J}_{z}^{3}-\left(3\widehat{X}+59\widehat{I}\right)\widehat{J}_{z}\right\}\left(\widehat{J}_{+}^{3}+\widehat{J}_{-}^{3}\right)\right.\\
&+\left.\left(\widehat{J}_{+}^{3}+\widehat{J}_{-}^{3}\right)\left\{11\widehat{J}_{z}^{3}-\left(3\widehat{X}+59\widehat{I}\right)\widehat{J}_{z}\right\}\right]
\end{split}
\end{equation*}
\begin{equation*}
\widehat{\mathcal{O}}_{6}^{6} = \frac{1}{2}\left(\widehat{J}_{+}^{6}+\widehat{J}_{-}^{6}\right)
\end{equation*}

\section{Crystal field parameters litterature review}
\label{appendixII}

In the pyrochlore literature, there exists mainly two ways to write the crystal field (CF) hamiltonian of the rare earth element: with Stevens equivalent operators $\widehat{\mathcal{O}}_{n}^{m}$ as in this study and in references \cite{Bertin2012,Klekovkina2014,Zhang2014}, and with the Wybourne operators \cite{Racah1942,Wybourne1965} $\widehat{\mathcal{C}}_{m}^{n}$, as in references \cite{Gingras2000,Mirebeau2007,Princep2015,Ruminy2016}. The Wybourne operators are defined as
\begin{equation}
\widehat{\mathcal{C}}_{m}^{n} = \sqrt{\frac{4\pi}{2n+1}}\widehat{Y}_{n}^{m}
\end{equation}
where $\widehat{Y}_{n}^{m}$ are the spherical harmonics operators. The CF hamiltonian for the $D_{3d}$ point group relevant for the rare earth element in the  pyrochlore compounds is then
\begin{equation}
\begin{split}
\widehat{\mathcal{H}}_{CF}^{(Wy)} &= W_{0}^{2}\widehat{\mathcal{C}}_{0}^{2} + W_{0}^{4}\widehat{\mathcal{C}}_{0}^{4} + W_{3}^{4}\left(\widehat{\mathcal{C}}_{-3}^{4} - \widehat{\mathcal{C}}_{3}^{4}\right) + W_{0}^{6}\widehat{\mathcal{O}}_{0}^{6} \\
&+ W_{3}^{6}\left(\widehat{\mathcal{C}}_{-3}^{6} - \widehat{\mathcal{C}}_{3}^{6}\right) + W_{6}^{6}\left(\widehat{\mathcal{C}}_{-6}^{6} + \widehat{\mathcal{C}}_{6}^{6}\right)
\end{split}
\end{equation}
where $W_{m}^{n}$ are the Wybourne crystal field parameters. Note that, in the literature, these quantities are often denoted $B_{m}^{n}$, the same way (except with an index exchange) as the Stevens crystal field parameters $B_{n}^{m}$.  Here, we prefer a different notation to avoid confusion.

 The Stevens operators $\widehat{\widetilde{\mathcal{O}}_{n}^{m}}$ are then derived from the Wybourne operators \cite{Kassman1970} :
 \begin{equation}
\widehat{\widetilde{\mathcal{O}}_{n}^{m}} = \left(\lambda_{n}^{m}\right)^{-1}\left(\widehat{\mathcal{C}}_{-m}^{n} +(-1)^{m} \widehat{\mathcal{C}}_{m}^{n}\right)
\end{equation}
 where the proportionality factors $\lambda_{n}^{m}$ are reproduced in Table \ref{lambda_nm_parameters} for those which are involved in the CF hamiltonian of the studied pyrochlore. Then, within the Hilbert space restricted to the ground multiplet $J$, this Stevens operator can be expressed as a function of the associated Stevens equivalent operators $\widehat{\mathcal{O}}_{n}^{m}$ used in this study and reproduced in Appendix \ref{appendixI}:
\begin{equation}
\widehat{\widetilde{\mathcal{O}}_{n}^{m}} = \theta_{n}(J)\widehat{\mathcal{O}}_{n}^{m}
\end{equation}
where the matrix element $\theta_{n}(J)$ are tabulated for the ground multiplet of all trivalent $4f$ ions in references \cite{Stevens1952,Hutchings1964} and reproduced here for Tb$^{3+}$ in Table \ref{Reduced_matrix_element_Tb3+}). The relationship between the Stevens and Wybourne crystal field parameters is then
\begin{equation}
B_{n}^{m} = \lambda_{n}^{m}\theta_{n}(J)W_{q}^{m}.
\end{equation}

We can now compare the CF parameters obtained in different studies for \TTO/ (Table \ref{Bnm_parameters}).

\begin{table}
 \begin{ruledtabular}
 \begin{tabular} {cccccc}
 $\lambda_{2}^{0}$ & $\lambda_{4}^{0}$ & $\lambda_{4}^{3}$ & $\lambda_{6}^{0}$ & $\lambda_{6}^{3}$ & $\lambda_{6}^{6}$ \\
\hline \noalign{\vskip 1mm}
 $1/2$ & $1/8$ & $-\sqrt{35}/2$ & $1/16$ & $-\sqrt{105}/8$ & $\sqrt{231}/16$ \\
 \end{tabular}
\end{ruledtabular}
\caption{Values of $\lambda_{n}^{m}$ parameters involved in the CF hamiltonian of the studied pyrochlore. }
\label{lambda_nm_parameters}
\end{table}

\begin{table}
 \begin{ruledtabular}
 \begin{tabular} {ccc}
 $\theta_{2}$ & $\theta_{4}$ & $\theta_{6}$ \\
\hline  \noalign{\vskip 1mm}
 $-1/99$ & $2/16335$ & $-1/891891$ \\
 \end{tabular}
\end{ruledtabular}
\caption{Matrix element $\theta_{n}(J)$ for Tb$^{3+}$ ($J = 6$).}
\label{Reduced_matrix_element_Tb3+}
\end{table}

\begin{table*}
 \begin{ruledtabular}
 \begin{tabular} {ccccccc}
  & $B_{2}^{0}$ & $B_{4}^{0}$ & $B_{4}^{3}$ & $B_{6}^{0}$ & $B_{6}^{3}$ & $B_{6}^{6}$ \\
\noalign{\vskip 0.5mm} \hline \noalign{\vskip 1mm}
Reference \cite{Bertin2012} & $-0.34$ & $4.9\cdot 10^{-3}$ & $4.3\cdot 10^{-2}$ & $-7.9\cdot 10^{-6}$ & $1.3\cdot 10^{-4}$ & $-1.1\cdot 10^{-4}$ \\
Reference \cite{Klekovkina2014} & $-0.28$ & $5.0\cdot 10^{-3}$ & $3.4\cdot 10^{-2}$ & $-7.5\cdot 10^{-6}$ & $1.1\cdot 10^{-4}$ & $-1.2\cdot 10^{-4}$ \\
Reference \cite{Zhang2014} & $-0.73$ & $4.1\cdot 10^{-3}$ & $5.9\cdot 10^{-2}$ & $-12\cdot 10^{-6}$ & $-5.0\cdot 10^{-4}$ & $-8.5\cdot 10^{-4}$ \\
Reference \cite{Princep2015} & $-0.28$ & $5.7\cdot 10^{-3}$ & $4.6\cdot 10^{-2}$ & $-8.0\cdot 10^{-6}$ & $1.6\cdot 10^{-4}$ & $-1.3\cdot 10^{-4}$ \\
Reference \cite{Ruminy2016}* & $-0.28$ & $4.7\cdot 10^{-3}$ & $4.1\cdot 10^{-2}$ & $-4.5\cdot 10^{-6}$ & $1.2\cdot 10^{-4}$ & $-1.4\cdot 10^{-4}$ \\
Reference \cite{Ruminy2016}** & $-0.27$ & $5.6\cdot 10^{-3}$ & $3.9\cdot 10^{-2}$ & $-6.9\cdot 10^{-6}$ & $1.7\cdot 10^{-4}$ & $-1.4\cdot 10^{-4}$ \\
This work & $-0.26$ & $4.5\cdot 10^{-3}$ & $-4.1\cdot 10^{-2}$ & $-4.5\cdot 10^{-6}$ & $-1.2\cdot 10^{-4}$ & $-1.4\cdot 10^{-4}$ \\
 \end{tabular}
\end{ruledtabular}
\caption{CF $B_{n}^{m}$ parameters (in \,meV) refined in different recent studies, together with those used in this work. *LS-coupling scheme. **Intermediate coupling scheme.}
\label{Bnm_parameters}
\end{table*}

One can be surprised that the sign of our $B_{4}^{3}$ and $B_{6}^{3}$ parameters are different from those of most of the literature. However, as pointed out by Bertin et al. \cite{Bertin2012}, when the sign of these two parameters are exchanged, there is no effect on the hamiltonian eigenvalues. This property is only true without magnetic field, which is the case for all the previous neutrons and optical studies, but does not hold under an applied magnetic field. Indeed, we find  much better agreement between our experimental results and calculations with $B_{4}^{3}<0$ and $B_{6}^{3}>0$ ; the eigenenergies at zero magnetic field are strictly identical when changing the sign of this two parameters.

\section{Wavefunctions for crystal-field states}
\label{appendixIII}

In  the included tables \ref{Wavefunction_CF_H}, \ref{Wavefunction_Vib_D21} and \ref{Wavefunction_Vib_D22} are given the wave functions for the ground and first excited doublets, without and with the vibronic coupling.

\begin{table*}[h!]
 \begin{ruledtabular}
 \begin{tabular} {ccccc}
  & $\ket{\psi_{+}^{0}} (0.0)$ & $\ket{\psi_{-}^{0}} (0.0)$ & $\ket{\psi_{+}^{1}} (13.5)$ & $\ket{\psi_{-}^{1}} (13.5)$ \\
\hline  \noalign{\vskip 1mm}
 $\ket{6}$ &  &  & & \\
 $\ket{5}$ & $0.35$ & & $-0.89$ &\\
 $\ket{4}$ & & $-0.91$ & &  $-0.37$\\
 $\ket{3}$ & &  & & \\
 $\ket{2}$ & $0.18$ & & $-0.25$ & \\
 $\ket{1}$ & & $-0.13$ & & $-0.14$\\
 $\ket{0}$ & & & & \\
 $\ket{-1}$ & $-0.13$ & & $-0.11$ & \\
 $\ket{-2}$ & & $-0.18$ & & $0.25$\\
 $\ket{-3}$ &  & & & \\
 $\ket{-4}$ & $0.91$ & & $0.37$ &\\
 $\ket{-5}$ & & $0.35$ & & $-0.89$\\
 $\ket{-6}$ &  &  & & \\
 \end{tabular}
\end{ruledtabular}
\caption{Wavefunction of the ground and first excited doublets, obtained by diagonalization of the CF hamiltonian (equation \ref{H_CEF}) without vibronic coupling ($D_{2}^{1} = D_{2}^{2} = 0$). The value in brakets following the wavefunction name is its associated eigenenergy.}
\label{Wavefunction_CF_H}
\end{table*}

\begin{table*}[h!]
 \begin{ruledtabular}
 \begin{tabular} {ccccc}
  & $\ket{\psi_{1}} (0.0)$ & $\ket{\psi_{2}} (2.76)$ & $\ket{\psi_{3}} (13.8)$ & $\ket{\psi_{4}} (16.6)$ \\
\hline  \noalign{\vskip 1mm}
 $\ket{6}$ &  &  & & \\
 $\ket{5}$ & $-0.30$ & $-0.09-0.12i$ & $-0.14+0.64i$ &  $-0.13-0.59i$\\
 $\ket{4}$ & $-0.44-0.43i$ & $-0.09+0.67i$ & $-0.14+0.09i$ &  $-0.17+0.27i$\\
 $\ket{3}$ & $-0.01i$ &  & &$-0.02$ \\
 $\ket{2}$ & $-0.14$ & $-0.06-0.08i$ & $-0.04+0.19i$ & $-0.03-0.16i$\\
 $\ket{1}$ & $-0.06-0.06i$ & $-0.01+0.10i$ & $-0.06+0.04i$ & $-0.05+0.08i$\\
 $\ket{0}$ & &  & & \\
 $\ket{-1}$ & $0.08$ & $0.06+0.08i$ & $-0.02+0.07i$ & $-0.02-0.09i$\\
 $\ket{-2}$ & $-0.10-0.10i$ & $-0.01+0.11i$ & $0.16-0.10i$ & $0.09-0.14i$\\
 $\ket{-3}$ &  & $0.01$ & & $-0.02$\\
 $\ket{-4}$ & $-0.62$ & $-0.41-0.54$ & $0.04-0.17i$ & $0.07+0.31i$\\
 $\ket{-5}$ & $0.21+0.20i$ & $0.02-0.15i$ & $-0.55+0.36i$ & $-0.33+0.51i$\\
 $\ket{-6}$ &  &  & & \\
 \end{tabular}
\end{ruledtabular}
\caption{Wavefunction of the ground and first excited doublets, obtained by diagonalization of the CF hamiltonian (equation \ref{H_CEF}) with a vibronic coupling parameter $D_{2}^{1} = -9.1\cdot 10^{-3}$\,meV. The value in brakets following the wavefunction name is its associated eigenenergy. Only coefficients of wavefunctions $> 10^{-2}$ are shown.}
\label{Wavefunction_Vib_D21}
\end{table*}

\begin{table*}[h!]
 \begin{ruledtabular}
 \begin{tabular} {ccccc}
  & $\ket{\psi_{1}} (0.0)$ & $\ket{\psi_{2}} (1.07)$ & $\ket{\psi_{3}} (13.7)$ & $\ket{\psi_{4}} (14.3)$ \\
\hline  \noalign{\vskip 1mm}
 $\ket{6}$ &  &  & & \\
 $\ket{5}$ & $-0.23+0.12i$ & $0.14-0.19i$ & $-0.44-0.46i$ &  $-0.19+0.59i$\\
 $\ket{4}$ & $-0.19+0.61i$ & $0.10+0.64i$ & $-0.26-0.01i$ &  $-0.12-0.25i$\\
 $\ket{3}$ & &  & &$-0.01$ \\
 $\ket{2}$ & $-0.12+0.07i$ & $0.07-0.10i$ & $-0.12-0.13i$ & $-0.05+0.16i$\\
 $\ket{1}$ & $-0.03+0.08i$ & $0.01+0.09i$ & $-0.08$ & $-0.04-0.07i$\\
 $\ket{0}$ & &  & & \\
 $\ket{-1}$ & $0.08-0.04i$ & $-0.05+0.07i$ & $-0.05-0.06i$ & $-0.03+0.08i$\\
 $\ket{-2}$ & $-0.04+0.13i$ & $0.02+0.12i$ & $0.18$ & $0.07+0.15i$\\
 $\ket{-3}$ &  & & & $-0.01$\\
 $\ket{-4}$ & $-0.56+0.30i$ & $0.38-0.53i$ & $0.17+0.18i$ & $0.08-0.26i$\\
 $\ket{-5}$ & $0.07-0.25i$ & $-0.04-0.23i$ & $-0.63-0.03i$ & $-0.27-0.56i$\\
 $\ket{-6}$ &  &  & & \\
 \end{tabular}
\end{ruledtabular}
\caption{Wavefunction of the ground and first excited doublets, obtained by diagonalization of the CF hamiltonian (equation 5) with a vibronic coupling parameter $D_{2}^{2} = -9.1\cdot 10^{-3}$\,meV. The value in brakets following the wavefunction name is its associated eigenenergy. Only coefficients of wavefunctions $> 10^{-2}$ are shown.}
\label{Wavefunction_Vib_D22}
\end{table*}

\end{document}